# The de Hass-van Alphen quantum oscillations in a three-dimensional Dirac semimetal TiSb$_2$

Wei Xia[1,2,3], Xianbiao Shi[4,5], Yushu Wang[6], Wenna Ge[1], Hao Su[1,2], Qin Wang[6,9], Xia Wang[1,7], Na Yu[1,7], Zhiqiang Zou[1,7], Yufeng Hao[6,8*], Weiwei Zhao[4,5*] and Yanfeng Guo[1,*]

[1]School of Physical Science and Technology, ShanghaiTech University, Shanghai 201210, China

[2]Shanghai Institute of Optics and Fine Mechanics, Chinese Academy of Sciences, Shanghai 201800, China

[3]University of Chinese Academy of Sciences, Beijing 100049, China

[4]State Key Laboratory of Advanced Welding and Joining, Harbin Institute of Technology, Shenzhen 518055, China

[5]Flexible Printed Electronic Technology Center, Harbin Institute of Technology, Shenzhen 518055, China

[6]National Laboratory of Solid State Microstructures, College of Engineering and Applied Sciences, Jiangsu Key Laboratory of Artificial Functional Materials and Collaborative Innovation Center of Advanced Microstructures, Nanjing University, Nanjing, 210093, China

[7]Analytical Instrumentation Center, School of Physical Science and Technology, ShanghaiTech University, Shanghai 201210, China

[8]Haian Institute of New Technology, Nanjing University, Haian, 226600, China

[9]School of Physics and Microelectronics, Zhengzhou University, Zhengzhou, Henan, 450001

*Corresponding authors:
haoyufeng@nju.edu.cn,
wzhao@hit.edu.cn,
guoyf@shanghaitech.edu.cn.

**Abstract**

We have used the de Hass-van Alphen (dHvA) effect to investigate the Fermi surface of high-quality crystalline $TiSb_2$, which unveiled a nontrivial topologic nature by analyzing the dHvA quantum oscillations. Moreover, our analysis on the quantum oscillation frequencies associated with nonzero Berry phase when the magnetic field is parallel to both of the *ab*-plane and *c*-axis of $TiSb_2$ finds that the Fermi surface topology has a three-dimensional (3D) feature. The results are supported by the first-principle calculations which revealed a symmetry-protected Dirac point appeared along the Γ-Z high symmetry line near the Fermi level. On the (001) surface, the bulk Dirac points are found to project onto the $\bar{\Gamma}$ point with nontrivial surface states. Our finding will substantially enrich the family of 3D Dirac semimetals which are useful for topological applications.

Three-dimensional (3D) topological semimetals (TSMs), which are often viewed as a "3D graphene", offer potential applications in next-generation spintronics and quantum computing.[1-3] The bulk bands in TSMs form a linear crossing in the 3D momentum space at a discrete point or at a line, namely, Dirac or Weyl point,[4-11] or nodal-line,[12-16] around which the linear dispersion is associated with unusual low-energy excitations that mimic relativistic Dirac or Weyl fermions predicated in high-energy physics.[17,18] The bulk bands crossing in 3D Dirac semimetals (DSMs) hosting Dirac fermions is fourfold degenerate touching at the Dirac point (DP) near the Fermi energy level $E_F$, which is protected by both time-reversal and space-inversion symmetries.[4-7] Once either the time-reversal or the space-inversion symmetry is broken, the spin degenerate DP evolves into a pair of spin-split Weyl points (WPs) with twofold degeneracy, thus driving a DSM into a Weyl semimetal (WSM).[8-11] One important fingerprint of the DP and WP is the corresponding nontrivial topological surface states,[5,9-11,19] which can be detected by surface sensitive experimental tools. In TSMs, exotic physical properties would emerge as a result of the relativistic equation of electrons motion, such as the nontrivial Berry phase,[20] the large linear transverse magnetoresistance (MR),[21,22] and negative longitudinal MR caused by the chiral anomaly,[23,24] etc. These intriguing physical properties offering potential applications have intensified the search for the unique topological states of matter in various stoichiometric materials.

The binary alloy TiSb$_2$ crystallizes in the tetragonal CuAl$_2$-type crystal structure,[25] comprising Ti atoms coordinated by eight Sb atoms, thus forming a square antiprism. The antiprisms form columns along the [001] direction in the structure by sharing quadratic faces. In addition, they share edges which results in a linkage of neighboring columns in the (001) plane, seen by the schematic views in Figs. S1(a) and (b). TiSb$_2$ has drawn considerable interest primarily as a suitable negative electrode for next generation sodium/lithium ion batteries and hybrid capacitors,[26,27] owing to its excellent energy-to-power density ratio performance, while the electronic structure and related properties have remained nearly uninvestigated. By performing de Hass-van Alphen (dHvA) quantum oscillation measurements and first-principle calculations, we have demonstrated the 3D DSM state in TiSb$_2$.

The crystal growth, phase and quality examinations, and resistivity measurement are presented in the Supplementary Information (SI). Magnetization measurements under magnetic fields up to $\mu_0 H = 7$ T were performed in a commercial magnetic properties measurement system (MPMS-3) from Quantum Design. Details for the first-principle calculations are also described in SI. The longitudinal resistivity $\rho_{xx}$ presented in the SI displays typical behaviors as those also observed in many other topological metals/semimetals.[28-31]

The calculated electronic density of states (DOS) of TiSb$_2$ along with atomic projected density of states (PDOS) without considering the spin-orbit coupling (SOC), plotted in Fig. 1(a), shows finite DOS at E$_F$, consistent with the metallic behavior. The Ti-3$d$ and Sb-5$p$ orbitals are widely distributed over all energies below and above the E$_F$ and dominate the low-energy states. The modest hybridization between the Ti-3$d$ and the Sb-5$p$ states implies a covalent bonding between Ti and Sb atoms. The electronic band structure computed without SOC is presented in Fig. 1(b), which clearly illustrates that two band crossing points (BCPs) along the Γ-Z and X-P high symmetry lines exist near the $E_F$. Fat-band analysis shows that the BCP along the Γ-Z is dominated by Ti-$d_{xy}$ and Sb-$p_y$ orbitals, while the BCP along the X-P is dominated by Ti-$d_{z^2}$ and Sb-$p_y$ orbitals. When the SOC is considered, as shown in Fig. 1(c), the BCP along the X-P line is fully gapped with a gap size of about 35 meV. The enlarged view of the band structure in Fig. 1(d) shows that a BCP along the Γ-Z direction exists about 50 meV above the $E_F$. Along the Γ-Z direction, the little group is $C_{4v}$ for any ***k*** point. The two crossing bands belong to different irreducible representations, $\Delta_6$ and $\Delta_7$, thus indicating that the BCP is symmetry protected. Because of inversion and time-reversal symmetries, the two crossing bands are both doubly degenerate. As such the BCP is fourfold degenerate DP, locating at (0, 0, $k_z^D \approx \pm 0.31 \times \frac{2\pi}{c}$). Fig. 2(e) shows the band dispersion around the BCP, which demonstrates that the point has linear dispersion. The calculated surface states on the (001) surface whose Brillouin zone (BZ) perpendicular to the Γ-Z line is presented in Fig. 1(f). On this surface, the bulk DPs are projected onto the surface $\overline{\Gamma}$ point. Their corresponding nontrivial topological surface states are pointed out by white arrows and clearly visible. These results further confirm the topological nature of TiSb$_2$.

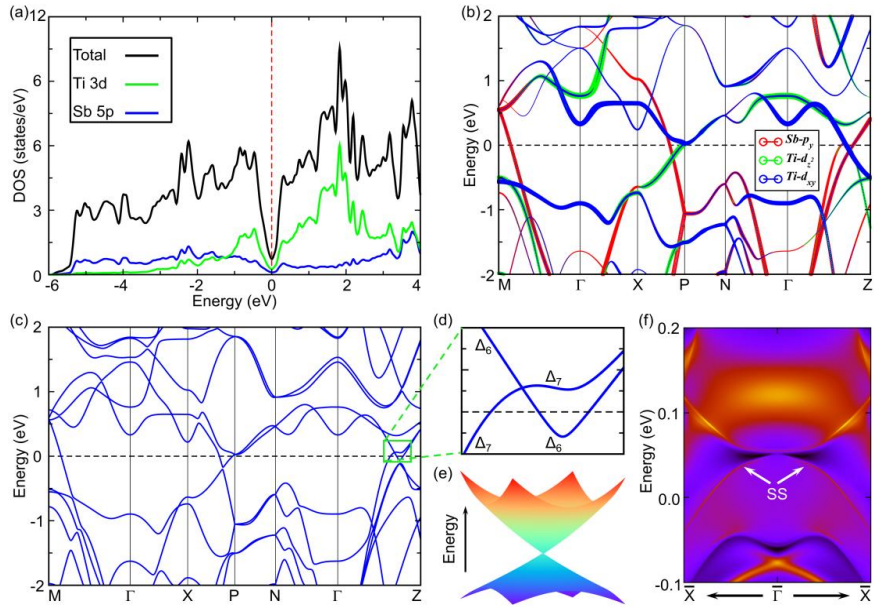

**Fig. 1.** (a) Total electronic density of states with projections on the orbitals for TiSb$_2$. Electronic band structures of TiSb$_2$ (b) without and (c) with SOC. The symbol size in (b) corresponds to the projected weight of Bloch states onto the Sb-$p_y$ (red), Ti-$d_{z^2}$ (green) and Ti-$d_{xy}$ (blue) orbits. The solid green area in (c) is enlarged in (d), where the DP is clear shown. (e) 2D band structure around the DP. (f) Surface and bulk band structure of TiSb$_2$ alone the $\overline{\Gamma}$-$\overline{X}$ direction on the (001) surface Brillouin zone. The nontrivial topological surface states stemming from the projection of bulk DP are pointed out by white arrows.

Since the quantum oscillation could not be detected in our MR measurements on TiSb$_2$ crystals, we alternatively used dHvA quantum oscillation measurements to probe the shape of Fermi surface (FS). The isothermal magnetizations $M(H)$ of TiSb$_2$ up to 7 T with $H//c$-axis and $H//ab$ plane at various temperatures were measured. As is shown in Fig. 2, $M(H)$ with both $H//c$ and $H//ab$ exhibit striking dHvA quantum oscillations at low temperatures with the oscillation magnitude reaching approximately 0.037 emu/g at the temperature ($T$) of 1.8 K, as is seen in Figs. 2(a) and (d), respectively. After a cautious subtraction of the diamagnetic background, remarkable periodic oscillations are visible in $\Delta M$ (= $M$ - $M_{\text{diamagnetic}}$) even up to 30 K ($H//c$) and 26 K ($H//ab$) as shown in Figs. 2(b) and (e), respectively.

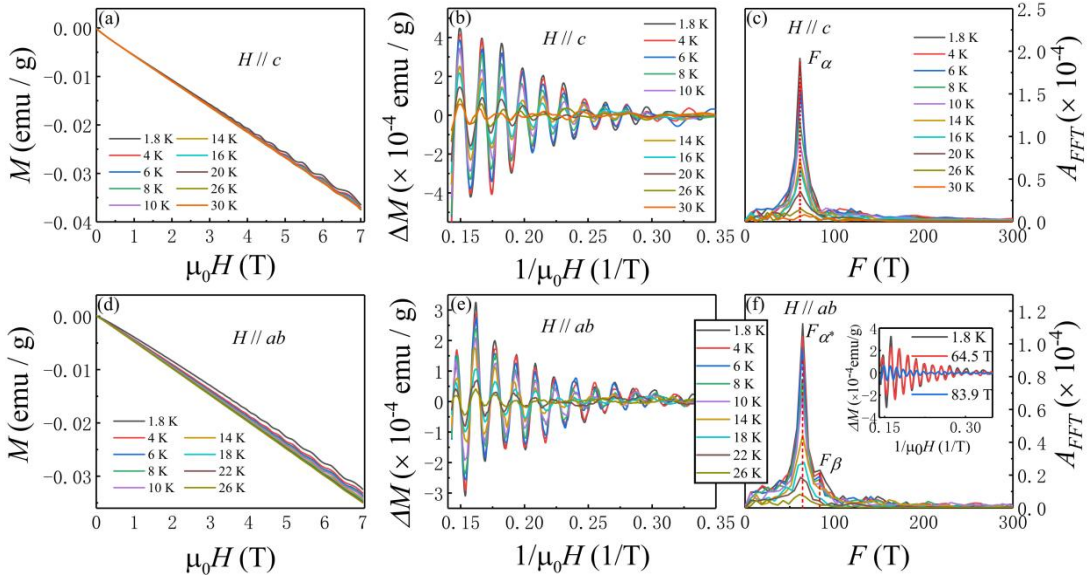

**Fig. 2.** (a) and (d) Magnetizations of TiSb$_2$ versus $1/\mu_0H$ for $H//c$ and $H//ab$ at various temperatures. (b) and (e) The quantum oscillations for $H//c$ and $H//ab$ versus $1/\mu_0H$ at various temperatures. (c) and (f) The Fast Fourier Transforms (*FFT*) spectra of *ΔM* at various temperatures. Inset in (f) shows the two filtered oscillatory parts of *ΔM*. The black line represents the raw dHvA oscillatory signal of TiSb$_2$ with $H//ab$.

TABLE I. Parameters derived from dHvA oscillations for TiSb$_2$.

| | $F$ (T) | $A$ (nm$^{-2}$) | $K_F$ (nm$^{-1}$) | $v_F$ (m/s) | $m^*$ /$m_e$ | $T_D$ (K) | $\tau_Q$ (s) | $\mu_Q$ (cm$^2$/Vs) | Berry phase |
|---|---|---|---|---|---|---|---|---|---|
| $H//c$ | 61.3 | 0.584 | 0.431 | 1.517×10$^6$ | 0.033 | 24.07 | 5.06×10$^{-14}$ | 2694.7 | 0.89$\pi$ |
| $H//ab$ | 64.5 | 0.614 | 0.442 | 1.65×10$^6$ | 0.031 | 37.7 | 3.23×10$^{-14}$ | 1832 | -0.66$\pi$ |
| | 83.9 | 0.8 | 0.504 | 2.01×10$^6$ | 0.029 | 55 | 2.21×10$^{-14}$ | 1340 | 1.01$\pi$ |

In Figs. 2 (b) and (e), the dHvA oscillations for both $H//c$ and $H//ab$ could be nicely fitted by the Lifshitz-Kosevich (L-K) formula[20]:

$$\Delta M \propto -B^\lambda R_T R_D R_s \sin[2\pi\left(\frac{F}{B} - \gamma - \delta\right)],$$

where $R_T = 2\pi^2 k_B T/\hbar\omega_c / \sinh\left(2\pi^2 k_B \frac{T}{\hbar\omega_c}\right)$, $R_D = \exp\left(-\frac{2\pi^2 k_B T_D}{\hbar\omega_c}\right)$, $R_S = \cos(\pi g m^*/2m_e)$, with $k_B$ being the Boltzmann constant, $\hbar$ being the Planck's constant, $F$ being the frequency of oscillation, $\omega_c = eB/m^*$ is the cyclotron frequency with $m^*$ denoting the effective cyclotron mass, and $T_D$ is the Dingle temperature defined by $T_D = \hbar/2\pi k_B \tau_Q$ with $\tau_Q$ being the quantum scattering lifetime. The exponent $\lambda$ is determined by the dimensionality of the material, where $\lambda$ = 1/2 and 0 are for the 3D and 2D cases, respectively. $R_T$ and $R_D$ are derived from the broadening in Landau levels induced by temperature effect in the Fermi-Dirac distribution and electron scattering, respectively. $R_T$ describes the temperature dependence of the oscillation amplitude, while $R_D$ captures the field dependent damping of the oscillation amplitude.

The fast Fourier transform (*FFT*) spectra of the dHvA oscillations are depicted in Figs. 2(c) and (f). For $H//c$, one strong frequency $F_\alpha$ = 61.3 T is obtained, which is closely related to the orthogonal extremal cross-sectional area (*A*) of the Fermi pockets by the Onsager relation $F = (\hbar/2\pi e)A$, thus giving $A_\alpha$ = 0.584 nm$^{-2}$. The effective cyclotron mass $m^*$ at E$_F$ can be estimated from the L-K fitting to the $R_T$, as is shown in Figs. 3(a) and (d), giving $m^*$ = 0.033 $m_e$ for $F_\alpha$, where $m_e$ denotes the free electron mass. The Fermi wave vector for $F_\alpha$ is estimated to be 0.431 nm$^{-1}$ from $k_F = \sqrt{2eF/\hbar}$ and the corresponding Fermi velocities $v_F = 1.52 \times 10^6$ m s$^{-1}$ as is calculated from $v_F = \hbar k_F/m^*$. Seen in Fig. 3(c), using the L-K formula to fit the field dependent amplitudes of the quantum oscillation at 1.8 K, the obtained Dingle temperature $T_D$ is 24.07 K and the corresponding quantum scattering lifetime $\tau_Q$ is $5.06 \times 10^{-14}$ s. Furthermore, the quantum mobility $\mu_Q$ of ~ $2.7 \times 10^3$ cm$^2$ /VS could be obtained from $\mu_Q = e\tau_Q/m^*$. For $H//ab$, a strong frequency $F_{\alpha^*} = 64.5$ T and a weak frequency $F_\beta = 83.9$ T are observed. Because the two frequencies are too close, we therefore applied a band pass filter to separate $F_{\alpha^*}$ and $F_\beta$ at 1.8 K. The results are shown by the inset of Fig. 2(f). The black and red lines represent the raw oscillatory and the oscillations of $F_{\alpha^*}$ and $F_\beta$, respectively. The parameters $A$, $m^*$, $k_F$, $v_F$, $T_D$, $\tau_Q$, $\mu_Q$ for both $F_{\alpha^*}$ and $F_\beta$ could be obtained by the similar analysis procedure as described above. The results are summarized in Table I to guide a comprehensive understanding about the FS.

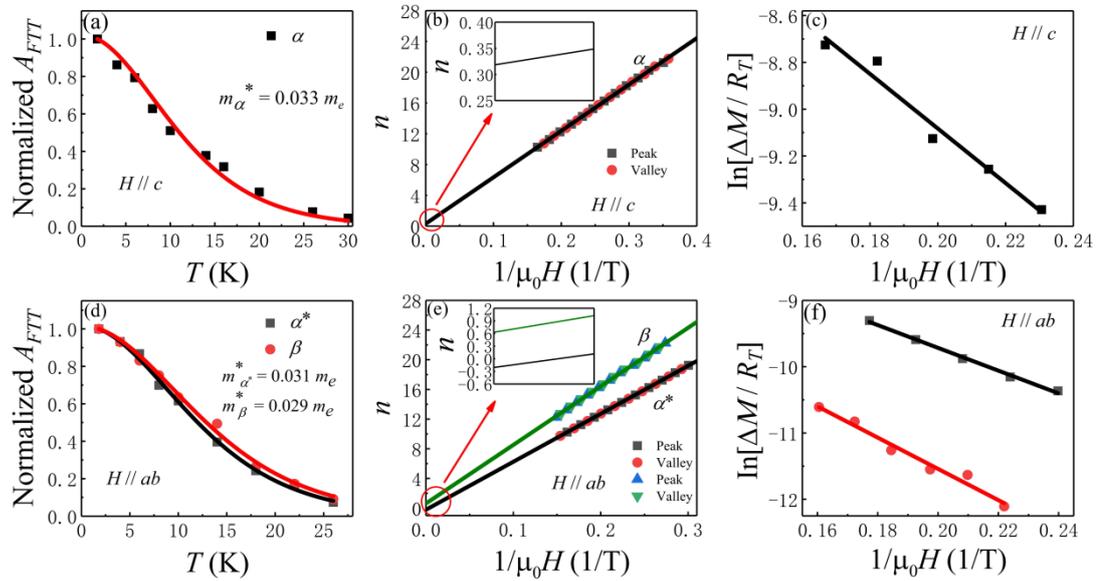

**Fig. 3.** (a) and (d) Temperature dependence of relative *FFT* amplitudes of the oscillations for $H//c$ and $H//ab$, respectively. (b) and (e) Landau indexes $N$-1/4 plotted against $1/\mu_0 H$ at 1.8 K for $H//c$ and $H//ab$, respectively. Inset in each enlarges the intercepts of the fitting. (c) and (f) Dingle plots of dHvA oscillations at $T = 1.8$ K for $H//c$ and $H//ab$, respectively, giving $T_{D\text{-}\alpha} = 24.07$ K in (c) and (f) $T_{D\text{-}\alpha^*} = 37.7$ K for $\alpha^*$ (the black line) and $T_{D\text{-}\beta} = 55$ K for $\beta$ (the red line).

The Landau level (LL) index fan diagram was constructed to examine the Berry phase of TiSb$_2$ accumulated along the cyclotron orbit, because the nontrivial Berry phase caused by the pseudo-spin rotation under a magnetic field is an key evidence for the Dirac fermions.[32,33] In the L-K equation, $\gamma$ ($\gamma = 1/2 - \phi_B/2\pi$) is the Onsager phase factor and $\delta$ represents the FS dimension-dependent correction to the phase shift, which is 0 for 2D system and ± 1/8 for 3D system in presence of nontrivial Berry phase. The LL index phase diagram is depicted in Figs. 3(b) and (e), in which the valley positions of $\Delta M$ correspond to the Landau indices of $N - 1/4$, and the peak positions of $\Delta M$ should correspond to the Landau indices of $N + 1/2 - 1/4$.[32] When $H//c$, the intercept derived from the linear extrapolation of the fit is 0.318, corresponding to a Berry phase $\phi_{B-\alpha}$ of $2\pi$ (0.318 + $\delta$). For a 3D FS, $\delta$ = ± 1/8, thus giving $\phi_{B-\alpha}$ = 0.89$\pi$, verifying a nonzero Berry phase. When $H//ab$, the intercept of $F_{\alpha^*}$ and $F_\beta$ are -0.2066 and 0.63, respectively, giving $\phi_{B-\alpha^*}$ = − 0.66$\pi$ and $\phi_{B-\beta}$ = 1.01$\pi$, indicating the nonzero Berry phase for both Fermi pockets. The analysis could be fully supported by the theoretical calculations.

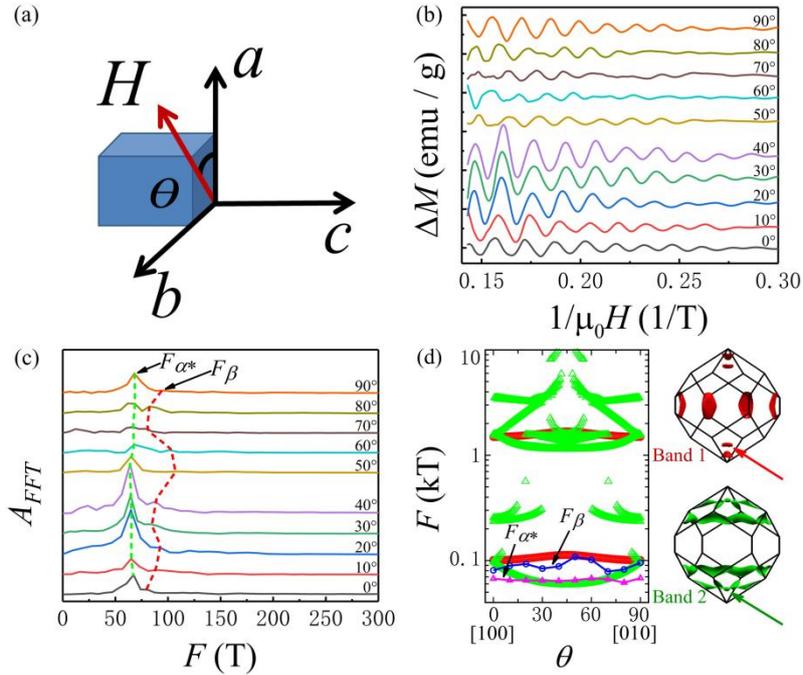

**Fig. 4.** (a) The schematic measurement configuration. (b) The quantum oscillation amplitudes

versus $1/\mu_0H$ at varied $\theta$ and $T = 1.8$ K. (c) The *FFT* spectra of *ΔM* at different $\theta$. (d) Left: the comparison between experimental and theoretical values of the fundamental frequencies with varied $\theta$. $F_{\alpha*}$ and $F_\beta$ represent the experimental values while thick circles (green) and triangles (red) denote the theoretical values. Right: the schematic view of electronic band structure producing the quantum oscillations, in which band 1 (red) yields the hole-pockets, and band 2 (green) leads to electron-pockets.

The angle dependent dHvA oscillations could provide further information about the shape of the FS. The schematic measurement configuration is shown in Fig. 4(a). The oscillations amplitudes versus $1/\mu_0H$ at various $\theta$ and $T = 1.8$ K are shown in Fig. 5(b). The $\theta$ dependence of fundamental frequencies is shown in Fig. 4(c). When the magnetic field was rotated between $B//a$ gradually to $B//b$, the experimentally determined $F_{\alpha*}$ and $F_\beta$ were compared with theoretical values, where the green thick circles denote the theoretical $F_{\alpha*}$ and the red triangles denote the theoretical $F_\beta$, as shown in Fig. 4(d). For $TiSb_2$, bands 1 (red) yields hole-pockets, while bands 2 (green) leads to electron-pockets. With a careful comparison between experimental and theoretical values, $F_{\alpha*}$ is assigned to the smallest electron-pockets pointed by the green arrow in bands 2, while $F_\beta$ is assigned to the smallest hole-pockets pointed by the red arrow in bands 1.

In summary, combing the de Hass-van Alphen effect measurements and first-principle calculations, we have demonstrated $TiSb_2$ as a three dimensional Dirac semimetal. Considering the promising use of $TiSb_2$ in next generation sodium/lithium ion batteries and hybrid capacitors, the revealed topological semimetal state would deepen the understanding about the electrical transport properties and hence enable the high efficient use of this material. Moreover, the Dirac point close to the Fermi energy level in $TiSb_2$ makes it very useful for topological applications.

The authors acknowledge the support by the Natural Science Foundation of Shanghai (Grant No. 17ZR1443300), the National Natural Science Foundation of China (Grant No. 11874264). W. Z. is supported by the Shenzhen Peacock Team Plan (Grant No. KQTD20170809110344233), and Bureau of Industry and Information Technology of Shenzhen through the Graphene Manufacturing Innovation Center (Grant No. 201901161514). Y.F.H acknowledges Distinguished Young Scholars Fund of Jiangsu Province (BK20180003), the National Key R&D Program of China (Grant No. 2018YFA0305800), the National Natural Science Foundation of China (Grant No. 51772145) and JiangHai Talent program of Nantong.

# Supplementary Information

## 1. Crystal growth and basic characterizations

TiSb$_2$ crystals were grown by using the self-flux method, starting from Ti powder (99.9%, Aladdin) and Sb (99.999%, Aladdin) granules mixed in a molar ratio of 1: 10. The mixture was loaded into an alumina crucible and was heated in a furnace up to 1100 $^{\circ}$C within 12 hrs, kept at the temperature for 5 hrs, and then slowly cooled down to 750 $^{\circ}$C with slowly decreasing the temperature at a rate of 1.5 $^{\circ}$C/h. The assembly was immediately put into a high-speed centrifuge to separate the excess Sb. The picture for a typical crystal with a dimension of 1 $\times$ 0.9 $\times$ 1.5 mm$^3$ is shown in Fig. S1(c).

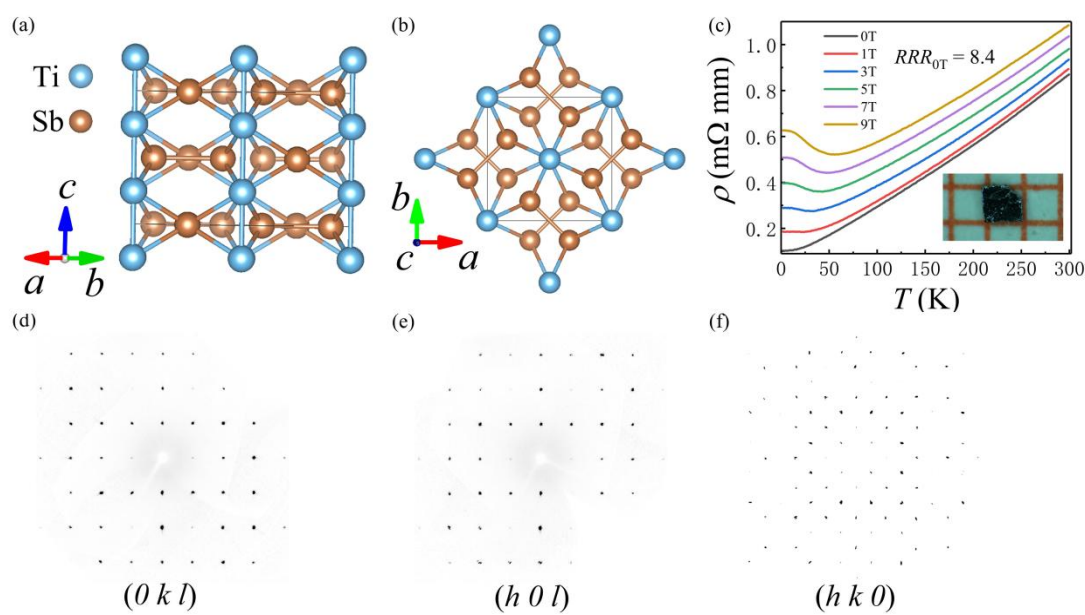

**Fig. S1.** (a) and (b) are the schematic views of the crystal structure of TiSb$_2$ from different orientations. (c) Temperature dependence of the longitudinal resistivity $\rho_{xx}$ at different magnetic fields. Inset shows a picture of a typical crystal. (d) - (f) Single crystal X-ray diffraction patterns in the reciprocal space along the (*0 k l*), (*h 0 l*), and (*h k 0*) directions.

The phase and quality examinations of TiSb$_2$ were performed on a Bruker D8 single crystal X-ray diffractometer (SXRD) with Mo $K_{\alpha1}$ ($\lambda$ = 0.71073 Å) at 298 K. The diffraction patterns could be well indexed by using a tetragonal unit cell in the space group *I*4/*mcm* (No. 140) with the lattice parameters $a$ = 6.649 Å, $b$ = 6.649 Å, $c$ = 5.7976 Å, $\alpha$ = 90°, $\beta$ = 90° and $\gamma$ = 90°, consistent with the parameters reported in other references.[1] The schematic structural views in Figs. S1(a) and (b) are drawn based on the refinement results of the SXRD data. The perfect reciprocal diffraction patterns without any other miscellaneous points, seen in Figs. S1(d) - (f), indicate a pure phase and high quality of the crystal. The resistivity measurements in a standard four-probe configuration were conducted in a Quantum Design DynaCool physical properties measurement system (PPMS). The temperature dependence of longitudinal resistivity $\rho_{xx}$ was measured with the magnetic field $B$ (= $\mu_0 H$) parallel to the *a*-axis and the electrical current $I$ along the *c*-axis. The $\rho_{xx}$ at $B$ = 0 T is presented in Fig. S1(c), displaying a metallic conduction upon cooling and yielding a residual resistance ratio (*RRR*) $\rho_{xx}$(300 K)/$\rho_{xx}$(2 K) of approximately 8.4. When $B$ is applied and gradually increased, $\rho_{xx}$ is apparently enhanced to be somewhat insulating with a clear plateau behavior at low temperature.

**2. First-principle calculations**

First-principle calculations were performed within the framework of the projector augmented wave (PAW) method[2,3] by employing the Perdew-Burke-Ernzerhof (PBE)[4] type generalized gradient approximation (GGA),[5] as encoded in the Vienna *ab* initio Simulation Package (VASP).[6-8] A kinetic energy cutoff of 520 eV and a Γ-centered *k* mesh of 12×12×12 were utilized in all calculations. During self-consistent convergence and structural relaxation, the energy and force difference criterion were defined as $10^{-6}$ eV and 0.01 eV/Å. The WANNIER90 package[9-11] was adopted to construct Wannier functions from the first-principle calculations results. The topological features of surface state spectra were investigate by using the WANNIERTOOLS code.[12] The SKEAF (Supercell K-space Extremal Area Finder) code[13] was used to extract the dHvA frequencies from the band structure calculations.